\documentclass[
aps,
prd,
10pt,
twocolumn,
superscriptaddress,
floatfix,
longbibliography,
nofootinbib
]{revtex4-2}

\usepackage{placeins}
\usepackage{amsmath,amssymb,mathtools,bm,bbm}
\usepackage{graphicx, color}
\usepackage{subfigure}
\usepackage[dvipsnames]{xcolor}
\usepackage{float}
\usepackage{multirow}
\usepackage{hyperref} %Automatically links \label and \ref commands; Always load last
\hypersetup{
    colorlinks=true,       % false: boxed links; true: colored links
    linkcolor=blue,        % color of internal links
    citecolor=blue,        % color of links to bibliography
    filecolor=magenta,     % color of file links
    urlcolor=blue,         % color of external links
    pdftitle={Resonant photon-axion conversion driven by axion dark matter oscillations},
    pdfauthor={Run-Min Yao, Xiao-Jun Bi, Peng-Fei Yin, Qing-Guo Huang}
}
\usepackage[utf8]{inputenc}
\usepackage[english]{babel}
\usepackage{orcidlink}

\graphicspath{{figures/}}

\begin{document}

% ---- Title ----
\title{Resonant photon-axion conversion driven by axion dark matter oscillations}
% ---- Authors / affiliations ----
\author{Run-Min Yao\orcidlink{0000-0002-7807-6713}}
\email{Contact author: yaorunmin@ucas.ac.cn}
\affiliation{School of Fundamental Physics and Mathematical Sciences, Hangzhou Institute for Advanced Study, UCAS, Hangzhou 310024, China}
\affiliation{University of Chinese Academy of Sciences, Beijing 100049, China}

\author{Xiao-Jun Bi\orcidlink{0000-0002-5334-9754}}
\email{bixj@ihep.ac.cn}
\affiliation{University of Chinese Academy of Sciences, Beijing 100049, China}
\affiliation{State Key Laboratory of Particle Astrophysics, Institute of High Energy Physics, Chinese Academy of Sciences, Beijing, China}

\author{Peng-Fei Yin\orcidlink{0000-0001-6514-5196}}
\email{yinpf@ihep.ac.cn}
\affiliation{State Key Laboratory of Particle Astrophysics, Institute of High Energy Physics, Chinese Academy of Sciences, Beijing, China}

\author{Qing-Guo Huang\orcidlink{0000-0003-1584-345X}}
\email{Contact author: huangqg@itp.ac.cn}
\affiliation{School of Fundamental Physics and Mathematical Sciences, Hangzhou Institute for Advanced Study, UCAS, Hangzhou 310024, China}
\affiliation{University of Chinese Academy of Sciences, Beijing 100049, China}
\affiliation{Institute of Theoretical Physics, Chinese Academy of Sciences, Beijing 100190, China}

%TC:ignore
% ---- Abstract ----
\begin{abstract}
Wave propagation in periodically time-dependent media can exhibit driven mode conversion that is absent in static or adiabatic descriptions. We show that coherent axion dark matter provides an intrinsic realization of this dynamics for photons propagating in a magnetic field. The background's periodic birefringence continuously rotates the photon polarization and thereby drives magnetic photon--axion mixing. Using Floquet theory, we derive resonant photon--axion conversion when integer harmonics of the axion oscillation frequency compensate the dispersion mismatch, $\Delta_\gamma-\Delta_a\approx n m_a$ with $n\in\mathbb{Z}$. This resonance enables efficient mixing far from conventional level crossing and is lost when the periodic drive is averaged out of the propagation Hamiltonian. The resulting conversion is bounded and unitary, distinct from parametric instability or axion decay. Our framework describes photon propagation in coherent oscillating axion backgrounds and predicts characteristic polarization signatures, which we illustrate in a blazar environment.
\end{abstract}

\maketitle
%TC:endignore
% ==========================================================
\noindent\textit{Introduction.}
Periodically driven media can support resonant mode conversion even when the corresponding static system is far from resonance. Floquet theory organizes this dynamics into harmonics of the drive and provides a standard description in condensed matter and quantum optics \cite{1965PhRv..138..979S,Sambe:1973cnm,2004PhR...390....1C,Oka:2009djg,Eckardt:2016lof}. A natural setting for such propagation is a coherent axion dark matter (DM) background. Axions and axion-like particles are well-motivated extensions of the Standard Model \cite{Peccei:1977hh,Peccei:1977ur,Weinberg:1977ma,Wilczek:1977pj,Cheng:1987gp} and candidates for DM \cite{Zwicky:1933gu,Bertone:2018krk,Preskill:1982cy,Abbott:1982af,Dine:1982ah,Arvanitaki:2009fg,Graham:2015ouw,Peccei:2006as,Marsh:2015xka,DiLuzio:2020wdo,Adams:2022pbo,OHare:2024nmr}. Their coherent oscillations make the optical properties of the background periodic, with a frequency set by the axion mass.

Two familiar consequences of the axion--photon interaction suggest different propagation pictures. In a static magnetic field, photons mix with axions, but a dispersion mismatch suppresses coherent conversion \cite{Raffelt:1987im,Raffelt:1996wa,Kuster:2008zz,Primakoff:1951iae,Sikivie:1983ip}. Conventional resonant conversion removes this mismatch at a level crossing, as in neutron-star magnetospheres where the plasma frequency equals the axion mass \cite{Pshirkov:2007st,Hook:2018iia}. A coherent axion background, meanwhile, rotates photon polarization \cite{Carroll:1989vb,Carroll:1998zi,Harari:1992ea} and can modify photon trajectories \cite{Plascencia:2017kca,Blas:2019qqp,McDonald:2019wou,Mcdonald:2020hjm}. Viewed separately, these effects describe conversion and background optics. Their coexistence motivates examining whether oscillating birefringence can itself drive conversion where static mixing is suppressed.

We demonstrate that periodic birefringence drives resonant photon--axion conversion through its noncommuting interplay with magnetic mixing: the background continuously changes the photon state that couples to the axion. Thus birefringence actively drives the conversion dynamics. For high-frequency photons propagating through a magnetized plasma, a co-rotating basis exposes Bessel-weighted harmonics of the mixing, with resonances at $\Delta_\gamma-\Delta_a\simeq n m_a$ ($n\in\mathbb{Z}$). The nonzero harmonics enable efficient conversion far from conventional level crossing, even with a static magnetic field. Floquet theory determines the harmonic couplings, resonance widths, and bounded conversion probabilities. The central result is the identification and quantitative description of this resonant interplay between background birefringence and photon--axion conversion.

In terms of phase matching, the periodic coupling compensates the phase mismatch accumulated between photon and axion amplitudes during propagation. Externally imposed magnetic modulation can also compensate a dispersion mismatch, through alternating-polarity magnets \cite{VanBibber:1987rq}, time-dependent fields \cite{Arias:2016zqu}, or spatial and temporal modulation underlying axion magnetic resonance \cite{Seong:2023ran}. Here the drive arises intrinsically from the axion background, which fixes its frequency and harmonic weights. The resonance locations and relative harmonic strengths therefore offer a potential probe of the axion mass and DM density, with the latter also dependent on the axion--photon coupling. Earlier studies with both an axion background and a magnetic field addressed mixed modes and polarimetry, as well as temporal-periodicity effects \cite{Espriu:2014lma}. Our focus is the resulting coherent, high-frequency forward-conversion regime in plasma and its polarization consequences.

Periodic axion backgrounds can also give rise to instabilities, in which electromagnetic perturbations grow by drawing energy from the background. Earlier work used Floquet theory to determine instability bands in the coupled axion--photon system with a magnetic field \cite{Masaki:2019ggg}, and studied parametric amplification in axion clumps \cite{Tkachev:2014dpa,Hertzberg:2018zte}, stimulated radio decay \cite{Caputo:2018ljp,Caputo:2018vmy}, and the axion-induced electromagnetic echo \cite{Arza:2019nta}. These photon-production processes differ from the unitary, bounded conversion of an incident beam described by our Hermitian forward-propagation Hamiltonian. We connect the harmonic resonances to conversion and circular polarization along an inhomogeneous line of sight, illustrating the effect in a blazar jet. The 3C~279 estimate is a source-dependent consistency example of the mechanism's observational consequences.

% ==========================================================
\noindent\textit{Driven photon-axion mixing.}
The axion may couple to electromagnetism through the operator $\mathcal{L}\supset-g_{a\gamma}aF\tilde{F}/4$, where $g_{a\gamma}$ is the photon--axion coupling  with units of $\mathrm{GeV}^{-1}$, $a$ is the axion field, and $F$ is the electromagnetic field tensor.
We consider photon propagation in the presence of a magnetic field and a coherently oscillating axion background. The axion-photon mixing equations in the presence of an external magnetic field can be reduced to a system of first-order differential equations via the WKB approximation, assuming the beam energy $\omega$ is much larger than the axion mass $m_a$. We consider photon propagation along the $z$-axis in the presence of a transverse magnetic field $\mathbf{B}_\mathrm{T}$.
In the linear polarization basis, the photon--axion system obeys the Schrödinger-like propagation equation,
\begin{equation}
    i \partial_z\Psi = \left[\mathcal{H}_0+\mathcal{H}_1\right]\Psi
\end{equation}
where the state vector $\Psi$ denotes the two photon polarization states and the axion field. The photon components correspond to linear polarizations oriented perpendicular and parallel to the direction of the external magnetic field. The dynamics is governed by a Hermitian mixing Hamiltonian with an explicit periodic time dependence.

The static part $\mathcal H_0$ describes conventional photon--axion mixing in a magnetized medium and is given by \cite{Raffelt:1987im,Raffelt:1996wa,Kuster:2008zz}
\begin{equation}
\mathcal{H}_0= -\begin{pmatrix} \Delta_\perp & 0 & 0\\ 0 & \Delta_\parallel & \Delta_\mathrm{B}\\ 0 & \Delta_\mathrm{B} & \Delta_a \end{pmatrix}.
\label{eq:mixing_matrix}
\end{equation}
Throughout this work, all dispersion and mixing effects are parametrized in terms of $\Delta$-parameters following the standard photon--axion mixing notation. Here, $\Delta_{\perp,\parallel}$ represent the photon dispersion for the two linear polarizations, accounting for both plasma effects and vacuum birefringence. In the systems considered here, vacuum QED effects are included in $\Delta_{\perp,\parallel}$ but remain subdominant to the plasma contribution, $\Delta_{\perp,\parallel} \simeq -\omega_\mathrm{pl}^2/(2\omega)$.  The effective plasma frequency is $\omega_\mathrm{pl} = \sqrt{4\pi \alpha n_e / m_e}$, where $\alpha$ is the fine-structure constant and $n_e$ and $m_e$ denote the electron number density and mass, respectively. The axion mass-induced dispersion is given by $\Delta_a = -m_a^2/(2\omega)$, while the photon--axion mixing is governed by $\Delta_{\rm B} = g_{a\gamma}B_\mathrm{T}/2$. The conventional plasma Faraday contribution is subleading for the effect of interest and can be neglected for the energy regime and the transverse magnetic field considered in this paper.

The axion background field is treated as classical and homogeneous,
\begin{equation}
    \bar{a}(t,z) = m_a^{-1}\sqrt{2\rho_\text{DM}}\cos(m_a t+\varphi_a)
\end{equation}
where $\rho_\text{DM}$ is the DM density,  and $\varphi_a$ is an a priori unknown phase.

In addition to the usual photon--axion mixing, the oscillating axion background induces a periodic birefringence for the photon, through the gradient coupling $g_{a\gamma}\partial_\mu\bar{a}$. In the linear polarization basis used here, the same effect appears as explicitly time-dependent terms in the photon subspace
\begin{equation}
\mathcal{H}_1(z)= -\begin{pmatrix} 0 & \Delta_F & 0\\ \Delta_F^* & 0 & 0\\ 0 & 0 & 0 \end{pmatrix},
\end{equation}
with
\begin{equation}
\Delta_F = i g_{a\gamma} (\partial_t{\bar{a}} + \partial_z \bar{a})/2.
\end{equation}
This birefringence term is purely imaginary, reflecting that it generates photon polarization rotation, analogous to but distinct from plasma Faraday rotation. We retain this periodic birefringence as the driving term $\Delta_F$ in the coupled propagation problem.

The driven contribution does not commute with the static mixing Hamiltonian, $[\mathcal{H}_0,\mathcal{H}_1(z)]\neq0$. Averaging the full time-ordered evolution operator therefore does not, in general, give the evolution generated by the averaged Hamiltonian, $\langle\mathcal{T}e^{-i\int\mathcal{H}(z)dz}\rangle\neq e^{-i\int\langle\mathcal{H}(z)\rangle dz}$, where the brackets denote an average over the background phase and $\mathcal{T}$ orders the Hamiltonians along the photon trajectory. Replacing $\Delta_F$ by its vanishing cycle average removes the periodic coupling from the propagation equation and eliminates the harmonic resonance derived below.

The Hamiltonian is manifestly Hermitian, and its only explicit time dependence originates from the oscillating axion background through the birefringence term,
\begin{equation}
\Delta_F = -i\Delta_\text{DM}\sin(m_a t+\varphi_a),
\end{equation}
where $\Delta_\text{DM} \equiv  g_{a\gamma}\sqrt{\rho_\text{DM}/2}$ sets the amplitude of the axion-induced drive. This term oscillates at frequency $m_a$ and acts as a coherent temporal drive of the photon--axion system. Along the photon trajectory we adopt the eikonal approximation $t\simeq z$ (with $c=1$), appropriate for forward-propagating waves with a well-defined carrier frequency. In this limit, the temporal oscillation of the axion background translates into a periodic modulation along the propagation direction with period $T=2\pi/m_a$.

The resulting propagation equation describes a three-level system subject to coherent periodic driving. A transformation to a co-rotating frame that follows the axion-induced phase modulation transfers the time dependence into the mixing terms. This allows the driven nature of the mixing to be made explicit while leaving the physical propagation amplitudes unchanged.

We therefore perform a time-dependent unitary transformation $\Psi(z)=U(z)\Phi(z)$ with $U=u\oplus1$, where $u$ acts on the photon subspace as
\begin{equation}
u(z) = \exp\left[-i\theta(z) \sigma_y\right],
\end{equation}
with $\sigma_y$ the second Pauli matrix.
The rotation angle $\theta(z)=\alpha\cos\phi(z)$, with $\alpha=\Delta_{\rm DM}/m_a$ and $\phi(z)=m_a z+\varphi_a$, is chosen such that the contribution from $U^\dagger i\partial_z U$ exactly compensates the rapidly oscillating birefringence term $\Delta_F$. This transformation is algebraically exact and merely redistributes the periodic dependence among the Hamiltonian matrix elements.

In the following we neglect QED vacuum birefringence, assuming $|\Delta_\perp-\Delta_\parallel|/2\ll m_a,\Delta_{\rm DM}$. This condition ensures that the two photon polarization states remain effectively degenerate on the scale of the axion-induced modulation, so that the driven mixing dynamics is not suppressed by polarization dephasing. We define $\Delta_\gamma\equiv(\Delta_\perp+\Delta_\parallel)/2$ as the average photon dispersion. Then the effective Hamiltonian becomes
\begin{equation}
\mathcal{H}'(z) = -
\begin{pmatrix}
\Delta_\gamma & 0 & \Delta_\mathrm{B}\sin\theta\\
0 & \Delta_\gamma & \Delta_\mathrm{B}\cos\theta\\
\Delta_\mathrm{B}\sin\theta & \Delta_\mathrm{B}\cos\theta & \Delta_a
\end{pmatrix}.
\label{eq:Hprime_simple}
\end{equation}
The unitary transformation leaves the physical propagation unchanged, but renders the periodic structure of the driven photon--axion mixing explicit, providing a natural starting point for the Floquet analysis below.

The full dynamics is a three-level Floquet problem, with the two photon modes coupled to the same axion through
$G^{(1)}=\Delta_{\rm B}\sin\theta(z)$ and $G^{(2)}=\Delta_{\rm B}\cos\theta(z)$. To identify when an effective two-level description applies, we first resolve these periodic couplings into harmonics.

The axion-induced phase modulation is periodic, so the couplings can be expanded in a discrete set of harmonics with amplitudes controlled by Bessel functions $J_\ell(\alpha)$. Explicitly, this follows from the standard Jacobi–Anger expansion,
\begin{equation}
e^{i\alpha\cos\phi}=\sum_{\ell\in\mathbb Z} i^\ell J_\ell(\alpha)e^{i\ell\phi}.
\end{equation}
Writing $\sin\theta$ and $\cos\theta$ in terms of $e^{\pm i\theta}$, the couplings can be decomposed into discrete Fourier harmonics,
\begin{equation}
    G(z)=\sum_{\ell\in\mathbb{Z}} g_\ell e^{i\ell m_a z}.
\end{equation}
Here $g_\ell^{(1)}$ ($g_\ell^{(2)}$) is nonzero only for odd (even) $\ell$, reflecting the parity structure of the two photon polarizations, and the magnitude of each harmonic scales as $g_\ell\sim\Delta_{\rm B}J_\ell(\alpha)$ up to phase factors. The parameter $\alpha$ controls the strength of the Floquet modulation, and multiple harmonics become relevant for $\alpha \gtrsim 1$ .

In this co-rotating basis, the driven photon--axion system naturally acquires a Floquet-harmonic structure \cite{Eckardt:2016lof}: the oscillating axion background provides a discrete ladder of harmonic driving channels separated by integer multiples of $m_a$. These harmonics shift the effective detuning between the dressed propagation modes by $\ell m_a$, so that a harmonic can become resonant even when the undriven modes are far from a level crossing.

For either photon--axion channel, we remove the free propagation phases and move to the interaction picture by defining $\Phi = \operatorname{diag}\left(e^{i\Delta_\gamma z},e^{i\Delta_a z}\right)\tilde \Phi$.
The $\ell$-th Fourier component of the coupling then acquires a phase factor
$e^{i\delta_\ell z}$, with detuning $\delta_\ell=\Delta_\gamma-\Delta_a-\ell m_a$.
Resonant mode mixing occurs when this phase is stationary, yielding the Floquet resonance condition
\begin{equation}
\Delta_\gamma-\Delta_a\approx n m_a,\quad n\in\mathbb{Z},
\label{eq:resonance_condition}
\end{equation}
corresponding to vanishing detuning $\delta_n$. In static mixing, the dispersion mismatch causes conversion amplitudes generated at different positions to accumulate different phases, limiting their coherent buildup. The $n$-th harmonic of the periodic coupling changes this relative phase evolution by offsetting the mismatch by $n m_a$. Near resonance, the residual phase varies slowly, allowing successive contributions to add coherently even when $\Delta_\gamma\neq\Delta_a$. Periodic magnetic-modulation schemes likewise restore coherent conversion by compensating the propagation-phase mismatch \cite{VanBibber:1987rq,Arias:2016zqu,Seong:2023ran}. In the present system, the modulation follows from axion-induced birefringence, so the available harmonics and their strengths are determined by the coherent background.

Near an isolated Floquet resonance labeled by $n$, retaining only the corresponding Fourier component constitutes a rotating-wave approximation, valid when the nonresonant harmonics in both photon channels satisfy $|\delta_\ell|\gg|g_\ell|$ for $\ell\neq n$. The odd or even parity of $n$ selects the photon mode coupled through $G^{(1)}$ or $G^{(2)}$, respectively. The other photon mode decouples at this order, leaving a two-level system undergoing coherent Rabi oscillations \cite{allen2012optical,scully1997quantum} with frequency
\begin{equation}
\Omega_{\rm R}=\sqrt{|g_n|^2+(\delta_n/2)^2}.
\end{equation}
The photon--axion conversion probability takes the standard form
\begin{equation}
P_{\gamma\to a} \approx \frac{|g_{n}|^2}{\Omega_{\rm R}^2} \sin^2\big(\Omega_{\rm R}\, z\big).
\end{equation}
Here $\gamma$ denotes the photon mode selected by the resonant harmonic in the co-rotating basis.
For fixed local parameters, the peak conversion probability attained over a Rabi oscillation is at least $50\%$ when $|\delta_n|\leq2|g_n|$. This interval defines the local mixing bandwidth, with full width $4|g_n|$ in detuning. Conversion during passage through a varying medium also depends on the accumulated propagation phase.

Removing the axion-induced drive, $\Delta_{\rm DM}=0$, recovers conventional photon--axion mixing. For weak modulation, $\Delta_{\rm DM}\ll m_a$, only the lowest harmonics contribute, and the driven resonance reduces to a small, single-frequency perturbation of the standard oscillation probability.

Additional derivations are provided in the Supplemental Material~\cite{SupplementalMaterial}\nocite{A:2025wjt,Dutton:2014xda,Espriu:2011vj,Ferrarese:2002ct,Navarro:1995iw,Schive:2014dra,Schive:2014hza}, where we also discuss an equivalent formulation in the circular polarization basis. In that representation, the axion background appears as a periodic modulation of the refractive indices, providing an alternative but equivalent perspective on the driven mixing.

This equivalent optical description also clarifies the relation to earlier axion-background studies. Ray evolution and polarization-dependent dispersion \cite{McDonald:2019wou}, mixed modes and polarimetry \cite{Espriu:2014lma}, and temporal instability bands \cite{Masaki:2019ggg} address related aspects of axion electrodynamics. To obtain the conversion considered here, the periodic optical response must be retained in the coupled evolution of photon and axion amplitudes. The near-resonant solution then determines how this response changes the buildup of conversion over a finite propagation distance.

The resulting resonance describes coherent transfer between forward photon and axion modes. Near an isolated resonance, their effective $2\times2$ Hamiltonian is Hermitian, and the probability derived above oscillates with propagation distance while remaining bounded. Periodic driving therefore enhances mode conversion without producing exponential growth of the propagating amplitudes.

An instability requires a different dynamical channel. In parametric amplification, the background couples positive- and negative-frequency sectors and can transfer energy into photon-pair production, as near $m_a\simeq2\omega$ \cite{Hertzberg:2018zte}. The forward, high-frequency approximation adopted here excludes this channel. The resonance condition instead matches the difference between photon and axion dispersion terms to a harmonic of the background, yielding oscillatory conversion of an incident beam. The two effects arise from periodic backgrounds but involve different modes and resonance conditions.

Throughout this treatment, the axion condensate is a prescribed coherent background, with recoil and depletion neglected. Its oscillation enters the propagation Hamiltonian through the Fourier coefficients $g_n$. The harmonic index thus labels Fourier components of the coherent drive, whose amplitudes determine the conversion strength in each channel. Further details are given in the Supplemental Material.

% ==========================================================
\noindent\textit{Representative propagation.}
To illustrate the physical consequences of Floquet resonances, we consider photon propagation in a medium with a slowly varying dispersion $\Delta_\gamma(z)$, in the presence of a coherent axion background over the relevant length scale.
This setup isolates the driven resonance mechanism without reliance on detailed source modeling.

The local detuning now changes along the trajectory. As the photon propagates, the dispersion mismatch can pass through the Floquet resonance condition $\Delta_\gamma(z)-\Delta_a\simeq n m_a$, activating resonant photon--axion mode conversion. Conversion is enhanced wherever the difference between the plasma and axion dispersion terms is compensated by a harmonic of the axion background, without requiring a crossing of the dispersion terms. A slowly varying profile can therefore bring the beam through successive resonances, with conversion building up coherently around each crossing.
\begin{figure}[!tbp]
\includegraphics[width=\columnwidth]{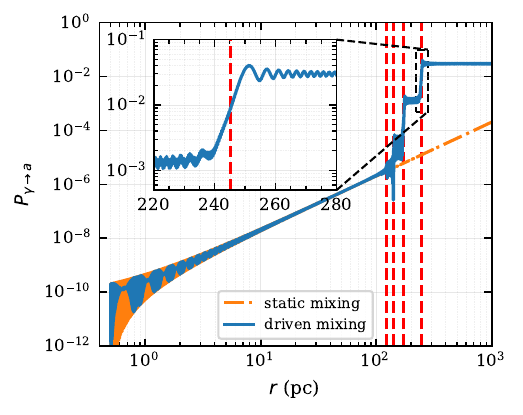}
\caption{Photon--axion conversion probability $P_{\gamma \to a}$ along a representative blazar jet.
The blue line shows the case of driven Floquet mixing with an oscillating axion background, while the orange line indicates the conventional adiabatic conversion obtained from the static mixing problem.
Vertical red dashed lines mark locations where $\Delta_\gamma(r) - \Delta_a \approx n m_a$, corresponding to $n = -1, -2, -3, -4$ from right to left under our negative plasma-dispersion convention.}
\label{fig:resonance}
\end{figure}

Figure~\ref{fig:resonance} shows the resulting photon--axion conversion probability $P_{\gamma\to a}$ as a function of propagation distance for representative parameters, specifically $\omega \sim \mathrm{eV}$, $m_a = 10^{-22}\,\mathrm{eV}$,
$g_{a\gamma}=5\times 10^{-11}\,\mathrm{GeV}^{-1}$, $\rho_\mathrm{DM}=1\,\mathrm{GeV\cdot cm^{-3}}$.
The driven Floquet case (blue) exhibits sharply enhanced conversion when the dispersion mismatch crosses a resonance band, whereas standard static mixing (orange) remains oscillatory and strongly suppressed. The vertical dashed lines indicate the locations where $\Delta_\gamma(z)-\Delta_a\simeq n m_a$, where the electron density $n_e\sim 0.1\,\mathrm{cm^{-3}}$ (corresponding to $\omega_\mathrm{pl}\sim 10^{-11}\,\mathrm{eV}$) and the magnetic field is $\sim 1\,\mathrm{mG}$ at the resonance position $\sim 10^2\,\mathrm{pc}$. Since $|\Delta_\gamma(r)|\gg|\Delta_a|$, the resonances occur where $|\Delta_\gamma(r)|\simeq |n|m_a$.

The local Rabi expression characterizes mixing at fixed parameters. During resonance passage, the conversion amplitude accumulates with the phase integral of the varying detuning. For an isolated resonance with locally linear detuning, define $\beta_n=|\partial_r\delta_n|_{r_{\rm res}}$. In the weak-coupling regime $|g_n|^2/\beta_n\ll1$, phase accumulation gives a characteristic conversion formation length $L_{\rm form}\sim\beta_n^{-1/2}$. This length can exceed the spatial interval defined by the local half-maximum mixing bandwidth, as derived in the Supplemental Material.

For the dominant $n=-1$ resonance in Fig.~\ref{fig:resonance}, $L_{\rm form}\sim3\,\mathrm{pc}$ is much smaller than the jet scale ($\sim\mathrm{kpc}$). Coherent buildup requires the DM field to remain coherent across this formation region. Its spatial coherence length ($\sim64\,\mathrm{pc}$ for a virial velocity $v\sim10^{-3}$) exceeds $L_{\rm form}$, and the photon crossing time ($\sim9\,\mathrm{yr}$) is much shorter than the DM coherence time ($\sim2\times10^5\,\mathrm{yr}$), supporting the coherent Floquet treatment of this passage.
We have verified that the resonance persists under moderate variations of $\Delta_\gamma(z)$ and in the presence of stochastic fluctuations, provided that the axion field remains coherent across the resonance region. These tests support persistence within the sampled environmental variations; more rapid variations that disrupt coherent accumulation across the layer can suppress the enhancement. Further numerical details are presented in the Supplemental Material.

The same resonance crossings also affect polarization. In the circular basis, the axion-induced modulation enters with opposite signs for the two photon states, making their conversion histories sensitive to the background phase and the propagation environment. This can generate circular polarization even from initially unpolarized radiation. For a given source, the axion phase is definite but observationally unknown and remains effectively fixed over the observations within the adopted coherence regime. The example below therefore assumes a uniform distribution over possible phases and uses the resulting spread in predicted circular polarization for an illustrative statistical consistency analysis.

% ==========================================================
\noindent\textit{Blazar 3C~279 as an illustration.}
We validate this theoretical framework with full numerical solutions and demonstrate that the driven mixing mechanism can imprint characteristic signatures on photon polarization during propagation.
As a concrete realization, we apply the formalism to optical polarimetry of the blazar 3C~279, with source-specific modeling details provided in the Supplemental Material.

To illustrate the potential observational impact of the mechanism, we adopt an illustrative statistical consistency criterion based on the spread of circular-polarization predictions over possible axion phases. Specifically, we require that the standard deviation of the axion-induced circular polarization, $\sigma_{\mathrm{std}}$, does not exceed the characteristic observational uncertainty of a given source. For 3C~279, we take $\sigma_{\Pi_C} \approx 0.30\%$ \cite{Liodakis:2021els}. For each fixed phase $\varphi_a$, the propagation model predicts a definite polarization. We assume a uniform distribution over the observationally unknown phase, obtaining a distribution of predictions with vanishing mean and nonzero variance. This comparison assesses the spread of predictions under the assumed phase distribution. Assigning a confidence level would require a statistical treatment of the source data and the unknown phase.

\begin{figure}[!tbp]
\includegraphics[width=\columnwidth]{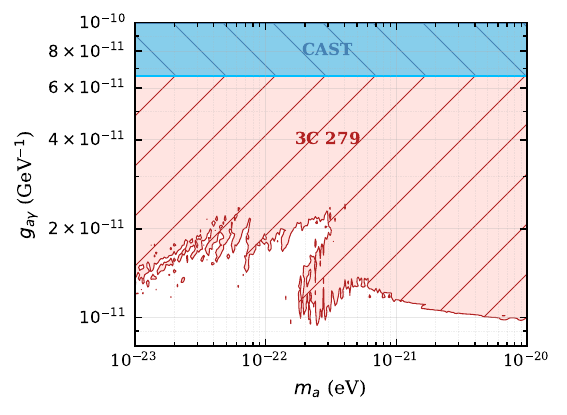}
\caption{Illustrative consistency contour for the axion--photon coupling $g_{a\gamma}$ derived from 3C 279 optical polarimetry. The shaded region indicates parameters for which the standard deviation of predicted circular polarization over an assumed uniform axion-phase distribution exceeds the adopted observational uncertainty. The driven mixing mechanism gives rise to a distinctive band structure in parameter space, directly reflecting the discrete Floquet resonance condition $\Delta_\gamma - \Delta_a \approx n m_a$.}
\label{fig:limits}
\end{figure}

Applying this criterion yields an illustrative consistency contour for the axion--photon coupling, shown in Fig.~\ref{fig:limits} together with CAST constraints \cite{CAST:2017uph} for reference. For axion masses in the range $m_a \in [10^{-23}, 10^{-20}]\,\mathrm{eV}$, the contour reaches $g_{a\gamma} \lesssim 2 \times 10^{-11}\,\mathrm{GeV}^{-1}$ for the adopted source parameters.
Since $J_n(\alpha)$ is maximized for $\alpha=\mathcal{O}(1)$, the conversion is most efficient when $m_a \sim \Delta_{\mathrm{DM}}$. For the benchmark coupling $g_{a\gamma}$ and dark-matter density, this condition selects the resonance region around $m_a \sim 10^{-22}\,\mathrm{eV}$ in Fig.~\ref{fig:limits}.
While the numerical reach of the contour depends on the specific noise properties of 3C~279, its qualitative structure is set by the underlying driven mixing dynamics.

In particular, the banded pattern in parameter space directly reflects the discrete Floquet resonance condition associated with the periodic axion background. As $m_a$ varies, the dominant contribution shifts between different resonance orders $n$, leading to alternating regions of enhanced and suppressed conversion. Near resonance, the driven mixing efficiently generates circular polarization, whereas off-resonant mixing between bands is suppressed.
This characteristic structure is a direct consequence of the time-dependent Floquet driving and does not rely on fine-tuned phase alignment.

The contour provides an order-of-magnitude illustration within the adopted source model and phase-distribution assumption. It demonstrates that the driven mixing mechanism can lead to observable, mechanism-specific polarization signatures at realistic sensitivity levels. The band structure and phase-dependent circular polarization connect the driven conversion to observables, with their magnitude determined by the adopted propagation environment.

% ==========================================================
\noindent\textit{Conclusions.}
Coherent axion DM can drive resonant photon--axion conversion through its periodic birefringence in a magnetized plasma. Retaining this drive yields a Floquet description with Bessel-weighted harmonic couplings and the resonance condition $\Delta_\gamma-\Delta_a\simeq n m_a$. The nonzero harmonics enable bounded conversion away from conventional level crossing. By continuously rotating the photon polarization that participates in magnetic mixing, the background birefringence opens a resonant conversion channel that is missed when the two effects are treated separately. Its resonance scales and harmonic weights are tied to the axion background's mass, density, and coherence.

The forward-propagation regime studied here describes unitary mode conversion, distinct from photon-pair production and temporal instability. In a nonuniform medium, the changing dispersion mismatch brings the beam through harmonic resonances. Conversion accumulates across a finite formation region around each crossing, so its effectiveness depends on coherence across that region. Environmental variations that disrupt this accumulation can suppress the effect.

In the representative jet model, these crossings generate circular polarization with a definite prediction for each axion phase. To assess the possible observational consequences when that phase is unknown, the 3C~279 example assumes a uniform phase distribution and compares the resulting spread of predictions with polarimetric sensitivity. The contour depends on this statistical assumption as well as the adopted source model. The broader result is a connection between coherent background oscillations, resonant conversion, and polarization observables. The harmonic structure offers access to the axion mass and to the combination of coupling and DM density that sets the modulation strength, motivating further applications in environments where propagation and coherence can be constrained.

% ==========================================================
\noindent\textit{Acknowledgments.}
The authors would like to thank the computing cluster of the School of Fundamental Physics and Mathematical Sciences at Hangzhou Institute for Advanced Study, UCAS, for providing the computational resources used in this work.
This work is supported by the National Natural Science Foundation of China under Grants No. 12547110, 12475065, 12250010, 12447131, 12447105, and 12575113.

\noindent\textit{Data availability.} The numerical data and code that support the findings of this article are available from the corresponding authors upon reasonable request.

% Shared bibliography for the article and Supplemental Material.
\bibliography{ref}

% Supplemental Material follows in the same PDF with continuous page numbers.
\clearpage
\onecolumngrid
\pdfbookmark[0]{Supplemental Material}{supplement}
\begin{center}
  \textbf{\large Supplemental Material for Resonant photon--axion conversion\\ driven by axion dark matter oscillations}\\[.2cm]
  \vspace{0.05in}
  {Run-Min Yao, Xiao-Jun Bi, Peng-Fei Yin, and Qing-Guo Huang}
\end{center}

\twocolumngrid

%%%%%%%%%% Merge with supplemental materials %%%%%%%%%%
\setcounter{equation}{0}
\setcounter{figure}{0}
\setcounter{table}{0}
\setcounter{section}{0}
\renewcommand{\theequation}{S\arabic{equation}}
\renewcommand{\thefigure}{S\arabic{figure}}
\renewcommand{\thetable}{S\arabic{table}}
\renewcommand{\thesubfigure}{(\alph{subfigure})}
\makeatletter
\renewcommand{\p@subfigure}{\thefigure}
\makeatother
\renewcommand{\theHequation}{S\arabic{equation}}
\renewcommand{\theHfigure}{S\arabic{figure}}
\renewcommand{\theHtable}{S\arabic{table}}

\onecolumngrid
This Supplemental Material (SM) contains additional calculation and derivation in support of the results presented in this work. First, we give a full derivation of the Hamiltonian presented in the main text. Then, we present additional details regarding the derivation of the two-level systems. Next, we discuss the relation between our work and parametric instability. Finally, we present a numerical analysis and apply the formalism to the relativistic jet of blazar 3C 279.

\section{Photon--Axion Conversion in an Axion Background}
In this section, we provide the step-by-step derivation of the Hamiltonian presented in the main text.

The Lagrangian for the photon field $A_\mu$ and axion field $a$ can be written as
\begin{equation}
\mathcal{L}=-\frac{1}{4}F_{\mu\nu}F^{\mu\nu}-A_\mu j^\mu + \frac{1}{2}(\partial_\mu a)^2-\frac{1}{2}m_a^2a^2-\frac{1}{4}g_{a\gamma} aF_{\mu\nu}\tilde{F}^{\mu\nu},
\label{lagrangain}
\end{equation}
where $F_{\mu\nu}$ and $\tilde{F}^{\mu\nu}$ denote the electromagnetic field tensor and its dual tensor, respectively, $j^\mu=(\rho,\mathbf{J})$ is the electromagnetic current density, and $g_{a\gamma}$ is the axion--photon coupling.  Adopting the temporal gauge $A_{0}=A^{0}=0$ and fixing the residual gauge freedom by imposing the Coulomb (radiation) gauge $\boldsymbol{\nabla}\cdot\mathbf{A}=0$, the equations of motion are:
\begin{gather}
\partial_{t}^{2}\mathbf{A}-\nabla^{2} \mathbf{A}=\mathbf{J}+g_{a \gamma} \partial_{t}a \boldsymbol{\nabla} \times \mathbf{A}-g_{a \gamma} \boldsymbol{\nabla} a \times\partial_{t}\mathbf{A},\label{eq:eom1}\\
-\boldsymbol{\nabla}\cdot\partial_{t}\mathbf{A}=\rho-g_{a \gamma}\boldsymbol{\nabla}a\cdot\left(\boldsymbol{\nabla}\times\mathbf{A}\right),\label{eq:eom2}\\
\partial_{t}^{2}a-\nabla^{2}a+m_{a}^{2}a=-g_{a \gamma}\partial_{t}\mathbf{A}\cdot\left(\boldsymbol{\nabla}\times\mathbf{A}\right).
\label{eq:eom3}
\end{gather}
For a weakly magnetized cold plasma, the dielectric tensor is isotropic, $\boldsymbol{\epsilon}=(1-\frac{\omega_\mathrm{pl}^2}{\omega^2})\mathbb{I}$, where $\omega_\mathrm{pl}=\sqrt{4\pi\alpha n_e/m_e}$ is the plasma frequency and $\omega$ is the energy of the photon--axion system. Here, $\alpha$ is the fine-structure constant, $n_e$ and $m_e$ denote the electron number density and mass, respectively. In the radiation gauge, longitudinal plasma oscillations decouple from the propagating electromagnetic modes. In Fourier space, the induced plasma current therefore contributes an isotropic dispersion term $\mathbf{J}_\mathrm{ind}(\omega)=-\omega^2(\mathbb{I}-\boldsymbol{\epsilon})\mathbf{A}(\omega)=-\omega_\mathrm{pl}^2\mathbf{A}(\omega)$ to both photon polarizations, which can be absorbed into an effective photon mass.

We apply the linearized perturbation to the equation and write the fields into background and perturbation
\begin{equation}
\mathbf{A}=\bar{\mathbf{A}}+\delta\mathbf{A},\qquad a=\bar{a}+\delta a,
\end{equation}
where $\bar{a}$ and $\bar{\mathbf{B}}=\boldsymbol{\nabla} \times\bar{\mathbf{A}}$ are the background axion and magnetic fields. The perturbations satisfy the equations of motion:
\begin{gather}
\partial_{t}^{2}\delta\mathbf{A}-\nabla^{2} \delta\mathbf{A}+\omega_\mathrm{pl}^2\delta\mathbf{A}=g_{a \gamma} \bar{\mathbf{B}}\partial_{t}\delta a + g_{a \gamma} \partial_{t}\bar{a} \boldsymbol{\nabla} \times \delta\mathbf{A}-g_{a \gamma} \boldsymbol{\nabla} \bar{a} \times\partial_{t}\delta\mathbf{A},\label{eq:1th-eom1}\\
-\boldsymbol{\nabla}\cdot\partial_{t}\delta\mathbf{A}=\delta\rho-g_{a \gamma}\boldsymbol{\nabla}\bar{a}\cdot\left(\boldsymbol{\nabla}\times\delta\mathbf{A}\right)-g_{a \gamma}\boldsymbol{\nabla}\delta a\cdot\bar{\mathbf{B}}, \label{eq:1th-eom2}\\
\partial_{t}^{2}\delta a-\nabla^{2}\delta a+m_{a}^{2}\delta a=-g_{a \gamma}\bar{\mathbf{B}}\cdot\partial_{t}\delta\mathbf{A}.\label{eq:1th-eom3}
\end{gather}
For a magnetic field perpendicular to the photon propagation direction and an axion gradient parallel to it, the source terms in Eq.\eqref{eq:1th-eom2} vanish. The last two terms in  Eq.~\eqref{eq:1th-eom1} characterize the birefringence effect induced by the axion DM. The first terms on the RHS of Eq.~\eqref{eq:1th-eom1} and Eq.~\eqref{eq:1th-eom3} reflect the axion--photon couplings. The conventional magnetic-mixing problem is recovered when the axion-background gradient term is absent.

We choose the direction of the wave vector $\hat{\mathbf{k}}$ along the $z$-axis and adopt the plane-wave Ansatz
\begin{equation}
    \delta\mathbf{A}=-i\tilde{\mathbf{A}}(z)e^{-i\omega t},\qquad \delta a=\tilde{a}(z)e^{-i\omega t},
\end{equation}
where $\delta\mathbf{A}$ includes an additional factor of $-i$ compared to $\delta a$. This choice ensures that the axion--photon coupling terms in the mixing matrix remain real, consistent with the standard form of the equations of motion found in most literature. Subsequently, we derive
\begin{gather}
(\partial_z^2+\omega^2-\omega_\mathrm{pl}^2)\tilde{\mathbf{A}}=-g_{a \gamma} \bar{\mathbf{B}}\omega\tilde{a}-g_{a\gamma}(\partial_{t}\bar{a}\partial_z+i\omega\partial_z \bar{a})(\hat{z}\times\tilde{\mathbf{A}}),\\
(\partial_z^2+\omega^2-m_a^2)\tilde{a}=-g_{a \gamma} \bar{\mathbf{B}}\omega\tilde{\mathbf{A}}.
\end{gather}
The equations can be rearranged as
\begin{equation}
(\partial_z^2+\omega^2+2\omega\mathcal{M})\psi=0 ,
\label{eq:second_order_wave}
\end{equation}
where $\psi=(\tilde{A}_\perp,\tilde{A}_\parallel,\tilde{a})$, and $\tilde{A}_\perp$ and $\tilde{A}_\parallel$ are the amplitudes for perpendicular and parallel polarization modes, respectively. The mixing matrix is
\begin{equation}
\mathcal{M}(z)= \begin{pmatrix} \Delta_\perp & \Delta_F & 0\\ \Delta_F^* & \Delta_\parallel & \Delta_\mathrm{B}\\ 0 & \Delta_\mathrm{B} & \Delta_a \end{pmatrix}.
\end{equation}
Here, $\Delta_{\perp,\parallel}$ represent the photon dispersion for the two linear polarizations, accounting for both plasma effects and vacuum birefringence. In the systems considered here, vacuum QED effects are included in $\Delta_{\perp,\parallel}$ but remain subdominant to the plasma contribution, $\Delta_{\perp,\parallel} \simeq -\omega_\mathrm{pl}^2/(2\omega)$. The axion mass-induced dispersion is given by $\Delta_a = -m_a^2/(2\omega)$, while the photon--axion mixing is governed by $\Delta_{\rm B} = g_{a\gamma}B_\mathrm{T}/2$ with the transverse magnetic field $B_\mathrm{T}$. The $\Delta_F$ term originates from the axion-gradient-induced birefringence and is formally analogous to the Faraday rotation term. Explicitly, it is given by
\begin{equation}
    \Delta_F\equiv \frac{g_{a\gamma}}{2\omega}(\partial_{t}\bar{a}\partial_z+i\omega\partial_z \bar{a})
\end{equation}
where the operator $\partial_z$ in the first term acts on the photon field.
Within the WKB approximation, $-i\partial_z\psi\approx\omega\psi$, the operator $\partial_z$ acting on $\psi$ can be replaced by $i\omega$, yielding
\begin{equation}
    \Delta_F= \frac{i}{2}g_{a\gamma}(\partial_{t}\bar{a}+\partial_z \bar{a}).
\end{equation}
The second-order equations of motion can be expressed in first-order form
\begin{equation}
(i\partial_z+\omega+\mathcal{M})\psi=0.
\label{eq:first_order_wave}
\end{equation}
In the derivation above, we have neglected the absorption effects of photons, the contribution of the cosmic microwave background's energy density to the dispersion relation, the plasma Faraday effect, and the QED vacuum birefringence effect.

Assuming a small DM virial velocity $v_{\mathrm{DM}} \ll 1$, we model the background field as
\begin{equation}
    \bar{a}(t,z) = m_a^{-1}\sqrt{2\rho_\text{DM}}\cos(m_a t+\varphi_a),
\end{equation}
where $\rho_\text{DM}$ is the DM density and $\varphi_a$ is the phase of the axion background. This leads to an oscillating birefringence term:
\begin{equation}
\Delta_F = -i\Delta_\text{DM}\sin(m_a t+\varphi_a^\prime),
\end{equation}
where the amplitude is $\Delta_\text{DM} \equiv  g_{a\gamma}\sqrt{(1+\mathcal{F}^2)\rho_\text{DM}/2}$ and the phase is $\varphi_a^\prime=\varphi_a+\arctan\mathcal{F}$. The factor $\mathcal{F}(z)=-(2m_a)^{-1}d\ln\rho_\text{DM}/dz$ can be used to describe possible corrections from spatial DM gradients, but will be neglected in the following. For fuzzy DM, $\mathcal{F}(z)$ is less than the virial velocity of DM. Along the photon trajectory we take the eikonal propagation approximation $t\simeq z$ (with $c=1$).

For reference, the relevant scales can be estimated as
\begin{align}
\Delta_{\perp,\parallel} &\simeq -\frac{\omega_{\rm pl}^2}{2\omega}\approx-6.89\times10^{-22}\,\mathrm{eV}\left(\frac{n_e}{1\,\mathrm{cm}^{-3}}\right)\left(\frac{\omega}{1\,\mathrm{eV}}\right)^{-1}, \\
\Delta_a &= -\frac{m_a^2}{2\omega}\approx-5\times10^{-45}\,\mathrm{eV}\left(\frac{m_a}{10^{-22}\,\mathrm{eV}}\right)^2\left(\frac{\omega}{1\,\mathrm{eV}}\right)^{-1}, \\
\Delta_B &= \frac{1}{2}g_{a\gamma} B_\mathrm{T}\approx4.88\times10^{-22}\,\mathrm{eV}\left(\frac{g_{a\gamma}}{5\times10^{-11}\,\mathrm{GeV^{-1}}}\right)\left(\frac{B_\mathrm{T}}{1\,\mathrm{G}}\right), \\
\Delta_{\rm DM} &= g_{a\gamma}\sqrt{\rho_{\rm DM}/2}\approx9.80\times10^{-23}\,\mathrm{eV}\left(\frac{g_{a\gamma}}{5\times10^{-11}\,\mathrm{GeV^{-1}}}\right)\left(\frac{\rho_\mathrm{DM}}{1\,\mathrm{GeV\cdot cm^{-3}}}\right)^{1/2}.
\end{align}
For typical astrophysical parameters, $\Delta_a$ is often negligible compared to $\Delta_\gamma$, while $\Delta_{\rm DM}$ provides a small but coherent periodic modulation of the mixing. The reduced Compton wavelength of axion is $\lambda=1/m_a\approx6.5\times10^{-2}\,\mathrm{pc}\left(10^{-22}\,\mathrm{eV}/m_a\right)$.

\section{Effective Two-Level Systems}
Applying $\psi (z)=e^{i\omega z}\Psi(z)$, the reduced Schrödinger-like equation can be written as
\begin{equation}
i \frac{d}{d z}\Psi(z) = \mathcal{H}(z)\Psi(z),\quad \mathcal{H}(z)=-\mathcal{M}(z),
\end{equation}
where $\mathcal{M}$ is the mixing matrix
including the oscillating birefringence term.
To solve the dynamics analytically, we perform a time-dependent unitary transformation $\Psi(z)=U(z)\Phi(z)$ to eliminate the rapid oscillations in the photon sub-block. The effective Hamiltonian becomes
\begin{equation}
\mathcal{H}'(z) = U^\dagger\mathcal{H}U - i U^\dagger\frac{dU}{dz}.
\label{eq:Hprime_def}
\end{equation}

Define $\phi(z) \equiv m_a z + \varphi_a$, the photon sub-sector of $\mathcal{H}$ can be decomposed using Pauli matrices:
\begin{equation}
\mathcal{H}_{\gamma}(z) = -\Delta_{\gamma}\,\mathbb I_2 - \delta\Delta_{\gamma}\,\sigma_z - \Delta_\text{DM}\sin\phi\,\sigma_y,
\end{equation}
where $\Delta_{\gamma} = (\Delta_\perp+\Delta_\parallel)/2$ and $\delta\Delta_{\gamma} = (\Delta_\perp-\Delta_\parallel)/2$.
We introduce a time-dependent unitary rotation acting in the photon subspace
\begin{equation}
u(z) = \exp\!\big[-i\theta(z) \sigma_y\big], \quad \theta(z) = \alpha \cos\phi(z).
\label{eq:U_transform}
\end{equation}
The transformation matrix is
\begin{equation}
U(z)=\begin{pmatrix} \cos\theta & -\sin\theta & 0\\ \sin\theta & \cos\theta & 0\\ 0 & 0 & 1 \end{pmatrix}.
\end{equation}
The term $iu^\dagger\partial_z u$ can be calculated as
\begin{equation}
    iu^\dagger\partial_z u = -\alpha  m_a \sin\phi \; u^\dagger \sigma_y u = -\alpha  m_a \sin\phi \;\sigma_y,
\end{equation}
where the final step utilises the fact that $\sigma_y$ commutes with itself. To cancel the $-\Delta_\text{DM}\sin\phi\,\sigma_y$ term in the Hamiltonian, we set $\alpha = \Delta_\text{DM}/m_a$, yielding the effective Hamiltonian
\begin{equation}
\label{eq:Hprime}
   \mathcal H'(z)=-\begin{pmatrix}
\Delta_\gamma+\delta\Delta_\gamma\cos(2\theta) & -\delta\Delta_\gamma\sin(2\theta) & \Delta_B\sin\theta\\[6pt]
-\delta\Delta_\gamma\sin(2\theta) & \Delta_\gamma-\delta\Delta_\gamma\cos(2\theta) & \Delta_B\cos\theta\\[6pt]
\Delta_B\sin\theta & \Delta_B\cos\theta & \Delta_a
\end{pmatrix}.
\end{equation}

When $|\delta\Delta_\gamma|\ll m_a$ and $ \Delta_\mathrm{DM}$, the phase evolution rate due to QED birefringence becomes negligible compared to the DM modulation frequency. In this regime, the photon states are effectively degenerate for axions, allowing the QED contribution to be safely neglected. The transformed Hamiltonian simplifies to
\begin{equation}
\mathcal{H}'(z) = -
\begin{pmatrix}
\Delta_\gamma & 0 & \Delta_\mathrm{B}\sin\theta\\
0 & \Delta_\gamma & \Delta_\mathrm{B}\cos\theta\\
\Delta_\mathrm{B}\sin\theta & \Delta_\mathrm{B}\cos\theta & \Delta_a
\end{pmatrix}.
\end{equation}
The two photon modes couple to the same axion through $G^{(1)} = \Delta_\mathrm{B}\sin\theta$ and $G^{(2)} = \Delta_\mathrm{B}\cos\theta$. To identify when an effective two-level description applies, we first resolve these couplings into harmonics.
Expressing $\sin\theta$ and $\cos\theta$ via the Jacobi–Anger identity
\begin{equation}
e^{i\alpha\cos\phi} = \sum_{\ell=-\infty}^{\infty} i^\ell J_\ell(\alpha) e^{i\ell\phi},
\label{eq:jacobi_anger}
\end{equation}
we expand $G^{(1,2)}$ into Fourier harmonics:
\begin{align}
G^{(1)}(z) &= \frac{\Delta_\mathrm{B}}{2i} \sum_{\ell} [1-(-1)^\ell] i^\ell J_\ell(\alpha) e^{i\ell\phi}, \\
G^{(2)}(z) &= \frac{\Delta_\mathrm{B}}{2} \sum_{\ell} [1+(-1)^\ell] i^\ell J_\ell(\alpha) e^{i\ell\phi}.
\label{eq:G_expanded}
\end{align}
Thus, each photon mode couples to the axion through a series of sidebands with frequencies $\ell m_a$.  $G^{(1)}$ and $G^{(2)}$ carry odd and even harmonics, respectively. We define the relevant coupling as
\begin{equation}
    G(z)=\sum_{\ell\in\mathbb{Z}} g_\ell e^{i\ell m_a z},
\end{equation}
with
\begin{align}
g_\ell^{(1)} &= \frac{[1-(-1)^\ell] i^{\ell-1}  e^{i\ell\varphi_a}}{2} J_\ell(\alpha) \Delta_\mathrm{B}, \\
g_\ell^{(2)} &= \frac{[1+(-1)^\ell] i^\ell e^{i\ell\varphi_a}}{2}  J_\ell(\alpha) \Delta_\mathrm{B}.
\end{align}

Near an isolated resonance, the harmonic parity selects one photon mode. When the nonresonant harmonics are weak compared with their detunings, as specified below, the coupling of the other photon mode can be neglected. Denoting the selected photon mode by $\gamma$ and its coupling by $G(z)$, the reduced equations are:
\begin{gather}
-i\frac{d}{dz}\Phi_\gamma = \Delta_\gamma \Phi_\gamma + G(z) \Phi_a, \\
-i\frac{d}{dz}\Phi_a = G(z) \Phi_\gamma + \Delta_a \Phi_a.
\label{eq:two_level}
\end{gather}
Moving to the interaction picture via $\Phi_\gamma = \tilde\Phi_\gamma e^{i\Delta_\gamma z}$, $\Phi_a = \tilde\Phi_a e^{i\Delta_a z}$, we obtain
\begin{gather}
-i\frac{d}{dz}\tilde\Phi_\gamma = \sum_\ell g_\ell e^{-i\delta_\ell z}\tilde\Phi_a,\\
-i\frac{d}{dz}\tilde\Phi_a = \sum_\ell g_\ell^* e^{i\delta_\ell z}\tilde\Phi_\gamma,
\label{eq:interaction_pic}
\end{gather}
with
\begin{equation}
    \delta_\ell=\Delta_\gamma-\Delta_a-\ell m_a.
\end{equation}
Resonance occurs when the phase oscillation is slow, i.e., when the momentum mismatch is compensated by the axion mass harmonics. The rotating-wave approximation retains the resonant harmonic $\ell=n$ with $\delta_n\approx0$ and requires $|\delta_\ell|\gg|g_\ell|$ for all $\ell\neq n$ in both photon channels. Since $g_n^{(1)}$ vanishes for even $n$ and $g_n^{(2)}$ vanishes for odd $n$, only one photon mode couples to the axion at this order. The resulting two-level system undergoes Rabi oscillations with an effective Rabi frequency
\begin{equation}
\Omega_{\rm R} = \sqrt{|g_{n}|^2 + (\delta_{n}/2)^2}.
\end{equation}
The conversion probability is given by:
\begin{equation}
P_{\gamma\to a} = \frac{|g_{n}|^2}{\Omega_{\rm R}^2} \sin^2\big(\Omega_{\rm R}\, z\big).
\end{equation}

A naive reading of $\Delta_\gamma-\Delta_a \simeq n m_a$ might suggest an incoherent process involving absorption of $n$ individual axion quanta from the background, which would be strongly suppressed by a high power of the coupling constant. However, that is not the appropriate description in the present setup. The dark-matter axion field is treated as a macroscopically occupied coherent background. In this limit, the interaction generates a periodic \emph{$c$-number drive} in the photon--axion propagation Hamiltonian. The problem is then naturally organized into Floquet sectors labeled by an integer $n$, and the relevant harmonic weights come from the Floquet decomposition of the coherent drive rather than from an isolated $n$-th order perturbative scattering amplitude.
So the $n$-th channel is weighted by the corresponding Bessel harmonic $g_n\sim \Delta_B J_n(\alpha)$ with $\alpha=\Delta_{\mathrm{DM}}/m_a$, rather than a naive suppression proportional to $g_{a\gamma}^n$. This is also why the dominant channels are typically low-order harmonics: for $\alpha=\mathcal O(1)$, the first few Bessel weights can be appreciable, whereas channels with $n\gg \alpha$ are automatically suppressed.

Additionally, we discuss the choice of transformations for diagonalizing the Hamiltonian $\mathcal{H}$. In general, Floquet theory can be directly applied to the three-level system. However, to isolate the dominant resonant dynamics and obtain a transparent physical picture, we diagonalize the photon subspace and reduce the problem to effective two-level photon--axion systems. For this alternative choice of basis, we retain the convention $\Psi=U_C\Phi$, where $\Psi$ is the original linear-polarization state and $\Phi=(\Phi_+,\Phi_-,\Phi_a)^T$ denotes the state in the circular basis. The columns of $U_C$ are the circular-polarization basis vectors expressed in the original basis, giving
\begin{equation}
U_C=\frac{1}{\sqrt{2}}\begin{pmatrix} 1 & 1 & 0\\ -i & i & 0\\ 0 & 0 & \sqrt{2} \end{pmatrix}.
\end{equation}
Thus $\Phi_\pm=(\Psi_\perp\pm i\Psi_\parallel)/\sqrt{2}$ and $\Phi_a=\Psi_a$, with $U_C^\dagger U_C=\mathbb{I}_3$. Since $U_C$ is constant, the transformed Hamiltonian is $\mathcal{H}_C=U_C^\dagger\mathcal{H}U_C$, giving
\begin{equation}
\mathcal{H}_C = -\begin{pmatrix}
\Delta_+ & \delta\Delta_\gamma & \frac{i\Delta_\mathrm{B}}{\sqrt{2}} \\
\delta\Delta_\gamma & \Delta_- & -\frac{i\Delta_\mathrm{B}}{\sqrt{2}} \\
-\frac{i\Delta_\mathrm{B}}{\sqrt{2}} & \frac{i\Delta_\mathrm{B}}{\sqrt{2}} & \Delta_a
\end{pmatrix}
\end{equation}
where $\Delta_\pm=\Delta_\gamma \mp\Delta_\text{DM}\sin\phi$ represent the photon dispersion for the two circular polarizations. Neglecting magnetic mixing and QED birefringence, the modulated photon states can be written approximately as
\begin{align}
\Phi_\pm&\simeq\exp\left[i\int^z dz' (\Delta_\gamma \mp\Delta_\text{DM}\sin\phi(z'))\right]\widetilde{\Phi}_\pm\\
&=e^{i\Delta_\gamma z}e^{\pm i\alpha\cos\phi}\widetilde{\Phi}_\pm,
\end{align}
Here $\widetilde{\Phi}_\pm$ are constant amplitudes in the absence of magnetic mixing, with integration phases absorbed into their definitions; $\Delta_\gamma$ and $\alpha$ are taken constant in this local calculation. Restoring magnetic mixing couples these modulated photon states to the axion. Expanding the phase factors $e^{\pm i\alpha\cos\phi}$ in Fourier harmonics then recovers the same resonance condition and Rabi oscillations as in the co-rotating basis. The two transformations thus expose the same periodic coupling, with $U(z)$ following the polarization rotation and $U_C$ expressing it through the accumulated phases of the circular modes.

Axion-induced frequency modulation has also been studied by McDonald and Ventura \cite{McDonald:2019wou}. Their Hamiltonian-optics calculation follows photon rays in a dynamical axion background, yielding frequency shifts, time delays, and refraction, but does not include magnetic conversion into a propagating axion mode. In the circular polarization basis, the background appears as a modulation of the photon dispersion, naturally directing attention toward frequency modulation and optical properties. This emphasis can obscure its role as a resonant drive of photon--axion conversion. Although either basis yields the same resonance, identifying that mechanism requires following how the periodic photon phase modifies the coupled photon--axion amplitudes. Its harmonics can then compensate the dispersion mismatch and allow conversion amplitudes to accumulate coherently, revealing birefringence as a resonant drive of magnetic mixing.

Espriu and Renau \cite{Espriu:2014lma} approximate the oscillating background by a piecewise-constant periodic profile, derive momentum gaps without a magnetic field, and analyze mixed modes and polarimetric effects with a magnetic field. Their no-field gap condition identifies momenta for which the temporal Floquet quasifrequency has no real solution. These gaps therefore characterize instability in the photon spectrum, rather than resonant transfer between photon and axion modes; the distinction from bounded conversion is developed below.

\section{Coherent Mode Mixing vs. Parametric Instability}
In the literature on axion electrodynamics, it is well established that an oscillating axion background can induce a parametric instability in the electromagnetic field. This is often formulated using the Mathieu equation and interpreted as the stimulated decay of axions into photons. Superficially, this resembles the propagation equation, Eq.~\eqref{eq:second_order_wave}, utilized in this work, where the mixing matrix $\mathcal{M}(z)$ also contains a periodic contribution from the axion DM background. In this appendix, we clarify that these two phenomena, despite originating from the same Lagrangian, describe physically distinct regimes and yield qualitatively different results.

\subsection{Common Origin}
Both formalisms derive from the standard axion--photon Lagrangian Eq.~\eqref{lagrangain}.
Consider a classical, coherently oscillating axion background
\begin{equation}
a(t)\sim a_0\cos(m_a t).
\end{equation}
Maxwell's equations acquire an explicit time dependence, characterized by Eqs. \eqref{eq:1th-eom1} to \eqref{eq:1th-eom3}. However, the subsequent reduction of these equations depends critically on the physical boundary conditions and the kinematic regime of interest.
\subsection{Regime I: Parametric Instability (The Mathieu Limit)}
The analysis of axion-induced parametric instability typically focuses on the spontaneous production and amplification of photon modes from the vacuum (or a low-occupation state). In this regime, one must retain the full second-order time dynamics, including both positive- and negative-frequency components of the field momentum modes:
\begin{equation}
\mathbf{A}(\mathbf{x},t) \sim A_\pm(t)\,e^{i\mathbf{k}\cdot\mathbf{x}}, \qquad A_\pm(t)=c(t)e^{-i\omega t}+c^*(t)e^{+i\omega t}.
\end{equation}
For circular polarizations, the equation of motion reduces to a Mathieu equation:
\begin{equation}
\ddot{A}_\pm + \omega^2 \left[1 \pm \epsilon\cos(m_a t)\right]A_\pm = 0,
\end{equation}
where $\epsilon \sim g_{a\gamma} a_0 m_a/\omega$. The defining feature of this regime is that the periodic axion background modulates the effective frequency squared of the oscillator. This modulation couples the positive-frequency ($e^{-i\omega t}$) and negative-frequency ($e^{+i\omega t}$) modes. When $m_a \simeq 2\omega$, the system enters an instability band where the field amplitude grows exponentially. \footnote{The instability of the photon--axion system—considering both axion DM and magnetic fields but omitting the plasma background—is explored in Ref.~\cite{Masaki:2019ggg}, where an instability condition $\sqrt{k_\gamma^2+m_a^2}+k_\gamma=nm_a$ is identified.} This corresponds to the stimulated decay of the axion field into photon pairs, a genuine parametric instability where the background acts as an energy pump. In addition, the band-gap structure $k_n=nm_a/2$ identified in Refs.~\cite{Espriu:2011vj,Espriu:2014lma} corresponds to a spectral manifestation of the similar underlying Floquet-instability structure.
\subsection{Regime II: Coherent Mode Mixing (The Propagation Limit)}
In contrast, this work investigates the propagation of a pre-existing electromagnetic wave with a fixed carrier frequency $\omega$ and well-defined direction. Employing the slowly varying envelope approximation, we express the field as a slowly evolving amplitude modulating a rapid oscillatory phase. Crucially, this procedure effectively integrates out backward-propagating and negative-frequency modes. The resulting first-order dynamics are governed by Eq.~\eqref{eq:first_order_wave}. While the axion background still introduces a periodic modulation in the mixing matrix $\mathcal{M}(z)$, its physical role changes fundamentally:
\begin{itemize}
    \item The modulation acts as a linear, Hermitian coupling between forward-propagating modes (e.g., photon polarization states).
    \item It does \emph{not} couple positive- and negative-frequency solutions.
\end{itemize}
Consequently, the system does not exhibit Mathieu-type exponential instabilities.
Instead, it displays norm-preserving Floquet resonances—enhanced but bounded mode conversion—when the eigenvalue difference matches the modulation frequency: $\lambda_i - \lambda_j = n m_a$.

\subsection{Theoretical Synthesis}
The distinction between these two phenomena can be understood by analogy to quantum optical systems.
\begin{itemize}
    \item[] \textit{Parametric Instability} (Regime I) is analogous to a driven harmonic oscillator. A periodic modulation of the oscillator's parameters leads to exponential growth of the mode amplitude within instability bands. In quantum terms, this represents parametric amplification, or particle creation from the vacuum.
    \item[] \textit{Floquet Mixing} (Regime II) is analogous to a driven multi-level system, such as Rabi oscillations in atomic physics. The relevant degrees of freedom are discrete internal states, and the periodic background induces transitions among them. Probabilities oscillate but remain bounded (unitary evolution).
\end{itemize}

From a formal Floquet-theory perspective, both systems involve linear differential equations with periodic coefficients. However, the relevant Floquet structures have different physical meanings:
\begin{itemize}
\item In the Mathieu-type case (instability), the periodic drive couples positive- and negative-frequency sectors. The corresponding Floquet exponents can acquire real growth rates, signaling exponential instability.
\item In the forward-propagation case (mixing), the periodic axion background generates a discrete set of Fourier-harmonic mixing channels. Resonances occur when one of these harmonics restores longitudinal phase matching between dressed propagation modes, thereby maximizing the mixing angle and producing bounded Rabi-like conversion rather than exponential growth.
\end{itemize}

At the level of the full system, the oscillating axion background defines a coherent periodic structure. In the local DM rest frame, or neglecting virial spatial gradients at leading order, this background is approximately time periodic. The photon--axion perturbations can therefore exchange energy and momentum coherently with the macroscopic axion background. In a microscopic description, the missing four-momentum is carried by the background condensate. In the classical-background language, the same exchange appears as coupling to Fourier harmonics of the periodic field, while the recoil of the macroscopically occupied condensate is neglected.
Photon-pair amplification and forward mode conversion use different channels of this periodic system. In the latter, the index $n$ labels the harmonic mediating conversion between forward amplitudes, not a photon-pair instability band.

Thus, while both phenomena arise within the framework of periodically driven systems, the propagation formalism employed in this work isolates the physics of coherent state mixing, explicitly distinct from the particle-production physics of parametric instability.

\section{Numerical analysis}
% ==========================================================
To numerically verify the analytical framework, we solve the Eq. \eqref{eq:first_order_wave} incorporating the mixing matrix $\mathcal{M}$ for a concrete astrophysical environment.  We model the relativistic jet of blazar 3C 279, focusing on the unpolarized optical R-band photons ($E_\gamma \approx 1.9\,\mathrm{eV}$) with a representative parameter set: an axion mass $m_a = 10^{-22}\,\mathrm{eV}$ and a coupling constant $g_{a\gamma} = 5\times10^{-11}\,\mathrm{GeV}^{-1}$. The radiation zone, located $r_0 = 0.5\,\mathrm{pc}$ from the central engine, is characterized by a magnetic field $B_0 = 0.5\,\mathrm{G}$ and electron density $n_{e,0} = 5\times10^{3}\,\mathrm{cm}^{-3}$ \cite{A:2025wjt}. The jet is modeled as a conical outflow with a Doppler factor $\delta \simeq 13$ \cite{A:2025wjt}, assuming scaling profiles $B(r) \propto r^{-1}$ and $n_e(r) \propto r^{-2}$ over a total length $L_{\rm jet} = 1\,\mathrm{kpc}$. To capture the fine-grained evolution of the mixing process, the calculation is performed over domain sizes ranging from $5 \times 10^{-7}\,\mathrm{pc}$ to $10^{-3}\,\mathrm{pc}$.

To isolate and illustrate the resonance structures dictated by the plasma profile, we adopt a uniform DM density $\rho_{\rm DM} = 1\,\mathrm{GeV\cdot cm^{-3}}$ ($2.63\times10^7\,M_\odot\mathrm{\cdot kpc^{-3}}$), and obtain
the photon--axion conversion probability shown in Fig.~\ref{fig:resonance}. Figure~\ref{fig:resonance} demonstrates resonant conversion exactly at the locations predicted by Eq.~\eqref{eq:resonance_condition}, while the behavior reduces to the adiabatic limit away from resonance. This excellent agreement confirms that the Floquet analysis accurately describes the physics of the driven system.
The conversion probability exhibits a distinct step-like growth, indicating that resonant conversion occurs within narrow spatial windows where the phase-matching condition is satisfied. The resonance positions depend solely on the jet's plasma profile and the axion mass, independent of the DM density $\rho_{\rm DM}$.
Thus, while a realistic non-uniform DM halo would modulate the amplitude of each "step" according to local $\rho_{\rm DM}$ and $B$ values, the qualitative structure of multiple resonances remains unchanged.

To distinguish the local mixing bandwidth from the length over which conversion forms during resonance passage, define $\beta_n=|\partial_r\delta_n|_{r_{\rm res}}$. If the detuning varies linearly and $g_n$ is approximately constant across the local mixing region, the half-maximum condition $|\delta_n|\leq2|g_n|$ gives the spatial full width
\begin{equation}
L_{\rm mix}\simeq\frac{4|g_n|}{\beta_n}
\simeq\frac{|g_{a\gamma}B_0J_n(\alpha)|r_0}{|n|m_a}.
\label{eq:mixing_width}
\end{equation}
The second expression uses $B=B_0r_0/r$, $\Delta_\gamma\propto-r^{-2}$, and negligible $\Delta_a$, so that $\beta_n\simeq2|n|m_a/r_{\rm res}$ for $n\neq0$. This width characterizes the region where the peak Rabi conversion probability at fixed local parameters is at least $50\%$.

The conversion accumulated during passage depends on the phase evolution across that region and its surroundings. For increasing detuning, write $x=r-r_{\rm res}$ and $\delta_n\simeq\beta_n x$. In the weak-coupling regime $|g_n|^2/\beta_n\ll1$, the incident photon amplitude changes little during an isolated crossing. Integrating the interaction-picture equations then gives the leading conversion amplitude, up to a constant phase,
\begin{equation}
\mathcal A_{\gamma\to a}(x)\simeq i g_n^*\int_{-\infty}^{x} ds\,e^{i\beta_n s^2/2}.
\label{eq:formation_amplitude}
\end{equation}
Here $g_n$ is evaluated at the resonance and the photon is initially in the selected mode. The local linear profile and constant coupling are extended beyond the formation region to evaluate the integral. Contributions accumulate coherently near its stationary phase, while more distant contributions increasingly cancel, giving the characteristic formation length
\begin{equation}
L_{\rm form}\sim\beta_n^{-1/2}.
\label{eq:formation_length}
\end{equation}
This estimate requires the detuning slope and coupling to vary little over $L_{\rm form}$. It specifies a characteristic scale; a full width extracted from the oscillatory conversion curve depends on the chosen criterion. For slow, adiabatic passage with $|g_n|^2/\beta_n\gg1$, the formation length instead scales as $|g_n|/\beta_n$, of the same order as $L_{\rm mix}$.

For the dominant $n=-1$ resonance in Fig.~\ref{fig:resonance}, the numerical energy prescription $\omega=E_\gamma/\delta\simeq1.9/13\,\mathrm{eV}$ gives $r_{\rm res}\simeq243\,\mathrm{pc}$, $\beta_n\simeq0.129\,\mathrm{pc}^{-2}$, and $|g_n|\simeq0.0341\,\mathrm{pc}^{-1}$. Thus $|g_n|^2/\beta_n\simeq0.009$, $L_{\rm mix}\simeq1.06\,\mathrm{pc}$, and $L_{\rm form}\sim2.8\,\mathrm{pc}$. The formation scale accounts for the growth over a few parsecs in the numerical conversion curve. Its small ratio to $r_{\rm res}$ justifies the local expansion, while $L_{\rm form}\ll L_{\rm jet}\sim1\,\mathrm{kpc}$ shows that the conversion is localized within a narrow part of the jet.

Coherent buildup during this passage requires the axion field to remain coherent over the formation region. For $m_a=10^{-22}\,\mathrm{eV}$ and virial velocity $v\sim10^{-3}$, the spatial coherence length $\ell_{\rm coh}\sim(m_a v)^{-1}\simeq64\,\mathrm{pc}$ exceeds $L_{\rm form}$, and the coherence time $\tau_{\rm coh}\sim(m_a v^2)^{-1}\simeq2\times10^5\,\mathrm{yr}$ exceeds the characteristic transit time $t_{\rm cross}\sim L_{\rm form}\sim9\,\mathrm{yr}$. Both conditions support treating the background as a coherent drive during this resonant interaction.

\begin{figure}[htbp]
\subfigure[Fluctuations in electron density]{
\includegraphics[width=.48\columnwidth]{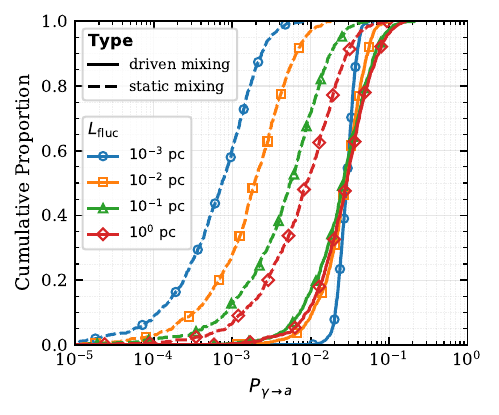}
\label{fig:robustness_ne}
}
\subfigure[Fluctuations in magnetic field]{
\includegraphics[width=.48\columnwidth]{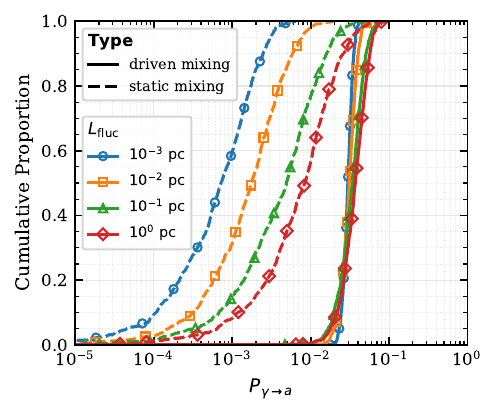}
\label{fig:robustness_B}
}
\caption{Empirical cumulative distribution function (ECDF) of the photon-to-axion conversion probability $P_{\gamma\to a}$. The colors represent different fluctuation correlation lengths $L_{\rm fluc}$, ranging from $10^{-3}$ to $1$ pc. Panels (a) and (b) correspond to stochastic fluctuations in the electron density and magnetic field, respectively.
Solid and dashed lines distinguish between driven Floquet mixing with oscillating DM background and adiabatic conversion obtained from the static mixing problem.
Each curve is generated from $N=1000$ random realizations. Note that the $x$-axis is on a logarithmic scale.}
\label{fig:robustness}
\end{figure}

In addition to the coherence of the axion drive, the conversion accumulated across a resonance layer depends on the plasma and magnetic-field profiles. Density perturbations change the local detuning, while magnetic-field perturbations change the mixing strength. We therefore examine their effects separately to determine whether the enhancement found for the smooth jet persists.

To assess the robustness of the resonance mechanism, we investigate the effects of stochastic fluctuations in the electron density and the magnetic field separately. The propagation path is partitioned into discrete domains of length $L_{\rm fluc}$. For each scenario, we introduce a random bias by scaling the baseline model value by a coefficient $\xi$, uniformly sampled from $[0.5, 1.5]$, thereby representing $50\%$ amplitude fluctuations. As shown in Fig. \ref{fig:robustness}, even with these 50\% amplitude variations, the mean conversion probability under Floquet-driven mixing remains consistently 1–2 orders of magnitude higher than in the standard static scenario across all considered correlation lengths. While small-scale fluctuations ($L_{\rm fluc} \ll L_{\rm form}$) are effectively self-averaged, larger-scale variations ($L_{\rm fluc} \sim 1$ pc) increase statistical variance by inducing local dephasing. Nevertheless, the empirical cumulative distribution function (ECDF) reveals that even the lower quantiles of these distributions significantly surpass the standard adiabatic case.

The static-mixing curves also increase with $L_{\rm fluc}$, although they remain well below the driven case. This behavior has a simple phase-matching interpretation: stochastic variations in the electron density or magnetic field contain spatial Fourier components, and those components can enhance conversion when their wave numbers compensate the photon--axion propagation mismatch. The same algebraic idea underlies quasi-phase-matching with spatially varying magnetic fields \cite{VanBibber:1987rq}, time-dependent magnetic-field enhancements \cite{Arias:2016zqu}, and axion magnetic resonance in externally modulated magnetic fields \cite{Seong:2023ran}.

Although these numerical tests do not establish a general robustness guarantee, they provide evidence that the resonant enhancement is robust to the sampled density and magnetic-field fluctuations. The degree of robustness depends on the fluctuation correlation length and on phase coherence across the resonance region.

Standard astrophysical emission mechanisms in blazars typically yield negligible circular polarization. However, the Floquet resonance breaks the degeneracy between left- and right-handed circular polarization states through the parity-violating coupling. This induces a non-zero net circular polarization, whose sign and magnitude depend on the definite but observationally unknown phase $\varphi_a$ of the axion background at the resonance onset.

This phase-dependent circular polarization arises from coherent, unitary mode mixing and is therefore qualitatively distinct from polarization generated by absorption or scattering processes.
Assuming a distribution over the unknown phase gives a distribution of polarization predictions. Its variance depends on that statistical assumption, the harmonic conversion strength, and the source and propagation model. For a single source, the phase is effectively fixed over the observations, and the ensemble describes our uncertainty in its value. We use the resulting spread of predictions in the illustrative consistency analysis below.

We apply the formalism to construct an illustrative consistency contour for the axion--photon coupling $g_{a\gamma}$ using optical polarimetry of 3C 279. The DM density is modeled using a fuzzy DM profile, which consists of a solitonic core and an outer NFW envelope \cite{Navarro:1995iw,Schive:2014dra} with the total mass $M_h\sim 10^{13} M_\odot$ \cite{Ferrarese:2002ct}. The core density profile is adopted as $\rho_c[1+0.091(r/r_c)^2]^{-8}$ with the core radius $r_c \simeq 1.6\,\mathrm{kpc} \left(m/10^{-22}\,\mathrm{eV}\right)^{-1} \left(M_{\rm h}/10^9\,M_\odot\right)^{-1/3}$ and density $\rho_c \simeq 2.9\times10^6\, M_\odot\,\mathrm{kpc}^{-3} (m/10^{-22}\,\mathrm{eV})^2 (M_{\rm h}/10^9\,M_\odot)^{4/3}$ \cite{Schive:2014hza}. The transition radius between the two regions is taken to be $r_t \simeq 3r_c$. The  concentration parameter of the NFW halo is determined using the concentration–mass relation at redshift $z=0.5$ \cite{Dutton:2014xda}. This configuration yields negligible spatial gradients with $\mathcal{F}\lesssim10^{-3}$ .

\begin{figure}[htbp]
\includegraphics[width=0.5\columnwidth]{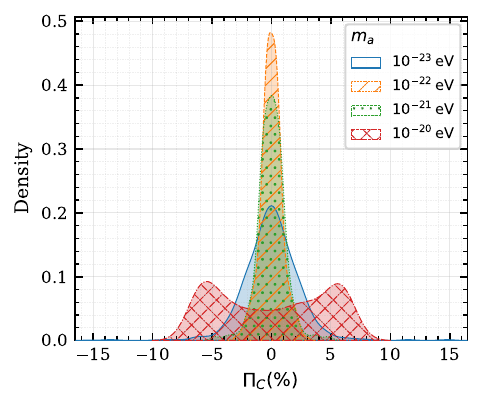}
\caption{Probability density of induced circular polarization for various axion masses $m_a$ (assuming $g_{a\gamma}=5\times10^{-11}\,\mathrm{GeV}^{-1}$), estimated via Kernel Density Estimation (KDE) by uniformly sampling the axion phase $\varphi_a\in[0,2\pi]$. The shaded profiles are generated using Gaussian kernels with Scott’s rule bandwidths. The transition from single-peaked to double-peaked distributions illustrates the varying sensitivity of the conversion process to the axion phase across different mass regimes.}
\label{fig:distribution}
\end{figure}

For each parameter pair $(m_a, g_{a\gamma})$, we solved the equations of motion 1,000 times using equally spaced values between $0$ and $2\pi$ for the axion phase $\varphi_a$ and assuming a $30\%$ initial linear polarization \cite{Liodakis:2021els}, thereby obtaining a well-resolved ensemble of induced circular polarization. We define the circular-polarization degree as $\Pi_C\equiv V/I$, where $V$ is Stokes circular polarization and $I$ is the surviving photon intensity. In Fig.~\ref{fig:distribution}, the resulting probability density of the degree of circular polarization $\Pi_C$ is visualized using kernel density estimation.

To construct an illustrative consistency contour, we require that the standard deviation of the induced circular polarization over a uniform initial-phase ensemble does not exceed the adopted observational uncertainty. For 3C~279 we use $\sigma_{\Pi_C}\approx0.30\%$ \cite{Liodakis:2021els}. The unknown phase $\varphi_a$ is assigned a uniform distribution for this analysis. The resulting polarization distribution can be bimodal, and its standard deviation summarizes the spread of predictions under that assumption. A confidence-level exclusion would additionally require a likelihood for the polarimetric data and a specified treatment of the unknown phase. Figure~\ref{fig:limits} shows the resulting source-dependent consistency contour, with CAST limits \cite{CAST:2017uph} included for scale comparison.

For $m_a \in [10^{-23}, 10^{-20}]\,\mathrm{eV}$, the consistency contour reaches $g_{a\gamma} \lesssim 2 \times 10^{-11}\,\mathrm{GeV}^{-1}$. The distinctive band-like structure of the resulting contour is a characteristic fingerprint of the Floquet-driven mode mixing mechanism, requiring driven, time-dependent mixing and the associated discrete resonance structure. Two generic features emerge from this mechanism:  (1) The conversion efficiency, affected by the Bessel functions $J_n(\Delta_\text{DM}/m_a)$, is maximized when the effective momentum mismatch is bridged by integer harmonics of $m_a$, a direct consequence of the periodic Floquet driving. The sharp dips in the consistency contour correspond to resonance “sweet spots” where this condition is optimally satisfied, with their locations and strengths depending on the adopted jet profile. (2) As $m_a$ varies, the dominant contribution shifts discretely between different resonance orders $n$, producing the characteristic banded pattern. The contour weakens between bands in off-resonance regimes, where the driven mixing becomes inefficient and circular polarization production is suppressed.

Although the numerical reach of the contour is normalized to the specific noise floor of 3C~279, its qualitative structure reflects a phenomenological consequence of the underlying mechanism, thereby illustrating the physical relevance of the framework.

\end{document}